\magnification=\magstep1
\font\foot=cmr9
\font\foots=cmsl9
\font\summ=cmr9
\font\nom=cmcsc10
\font\gr=cmss10 scaled \magstep1
\centerline{\gr Does light exert Abraham's force in a transparent medium?
\footnote{*}{\foot{Dedicated to Prof. Hans-J\"urgen Treder on his 70th birthday.}}
\footnote{\ddag}{\foot {Contributed to: {\foots From Newton to Einstein
(A Festschrift in Honour of the 70th Birthday of Hans-J\"urgen
Treder)}, W. Schr\"oder Editor, Bremen: Science Edition 1998.}}}
\par
\vbox to 0.4 cm{}
\centerline{{\nom S.~Antoci} and {\nom L.~Mihich}}
\centerline{\sl Dipartimento di Fisica ``A.~Volta''~-~Pavia, Italy}
\par
\vbox to 0.4 cm{}

\noindent{\sl Abstract:} {\summ Abraham's force seems to have been observed in
very low frequency experiments, but the existence of an Abraham's force exerted by
light is still to be proved. In fact, the only experiments performed with
light have measured the radiation pressure exerted at the interface between
different dielectric media, and their value for inferring conclusions about
the existence of Abraham's force is at best opinable. Twenty years ago Brevik
proposed an experiment for checking what forces light exert in the interior of a
transparent medium. It is argued that his proposal is feasible today with a
detection method now at hand, thanks to the great improvements achieved
in manufacturing optical fibres with the required physical properties.}\par
\vbox to 0.6 cm{}
{\gr 1. - Introduction}\par\bigskip
Let us figure out a physicist who has engaged in even a cursory survey of
the huge amount of theoretical literature accumulated in more than
one century about the forces that electromagnetism supposedly
produces in dielectric media: he has patiently compared the divergent
outcomes that eminent theoreticians have arrived at; he has doubtfully
pondered the relative advantages of the microscopic versus the macroscopic
approach; he has followed the alternating fortunes of electrostriction
and magnetostriction {\footnote{\foot $^{1)}$}{Accounted for already in the
seminal works by Helmholtz [1],[2], they have no place in the relativistic
proposals set forth by Minkowski [3] and Abraham [4], [5], but are present
(with a different shape) e.g. in the paper by Einstein and Laub [6].}};
he has wondered in disbelief why the supposedly universal tool of quantum
mechanics has been so inept, in a time span of seventy years, in shedding some
light on this issue. While challenged by such a confusing state of affairs,
he has even been summoned to surrender, and recognize the evident vanity
of the whole effort {\footnote{\foot $^{2)}$}{For instance, in Refs. [7]-[10]
it is argued, under natural assumptions like those introduced in Ref. [9], that
the way one splits the overall macroscopic energy tensor into a matter term and
a radiation term is just a convention, merely dictated by reasons of
convenience, and maybe of elegance.}}.\par
No wonder whether this physicist, like the simple Simon of the stinging
tale [11] written by J.L. Synge and entitled {\sl ``On the present status
of the electromagnetic energy tensor''}, might feel that his very
perception of an objective reality is at stake, and that, as an antidote,
he had better giving a look to whatever relevant information
experimentalists may have gathered on these forces, in the hope
that the latter have turned out to behave like classical, macroscopic
forces usually do, and that they have not been found to depend on the
mental act of splitting the overall energy tensor in two.
\par
\vbox to 0.8 cm{}

{\gr 2. The experimental evidence for Abraham's force}
\bigskip

Given the huge amount of theoretical literature, our natural philosopher
might expect to be confronted with an equally fair production on the
experimental side, but to his dismay, he will discover that all the papers
dealing with experimental findings can be easily contained in a small drawer
{\footnote{\foot $^{3)}$}{See, e.g., the accurate report [12] published by
Brevik in 1979; to our knowledge, just one further finding related to our
issue, about the angular momentum of radiation in dielectrics [13], has
occurred in the meantime.}}.\par
Clear evidence for the Abraham force {\footnote{\foot $^{4)}$}{A simple
derivation of Abraham's energy tensor, found by W. Gordon [14] in 1923,
is outlined in the Appendix.}} was retrieved with two kinds of experiments,
both availing of low frequency electromagnetic fields [15]-[17]. One experiment
performed by Walker and Lahoz will be summarized here, since it seems [12] to
have provided the most unambiguous outcome up to date.\par
To do so, let us first specialize the expression of Abraham's generalized
force density given in the Appendix for the case when a global coordinate
system exist in which the components of the metric are:

$$g_{ik}=\eta_{ik}\equiv{diag(1,1,1,-1)},\eqno(1)$$
and the components of the four-velocity of matter are

 $$u^1=u^2=u^3=0,\ u^4=1.\eqno(2)$$
Let us assume also the vanishing of the four-current ${\bf s}^i$; then
eq. (A33) reduces to:

$${\bf f}_{\rho}={-{{\epsilon\mu-1}\over \mu}
\left[F_{4\lambda}F_{\rho}^{~\lambda}\right]_{,  4}},~~{\bf f}_4=0,\eqno(3)$$
where $\epsilon$ and $\mu$ stand respectively for the dielectric constant
and for the permeability of a homogeneous and isotropic medium at rest.
Greek indices run from 1 to 3 and label the spatial components, while a comma
signals ordinary differentiation. Then in m.k.s. units Abraham's ordinary
force density takes the well known expression:

$${\bf{\vec f}}={{n^2-1}\over{c^2}}{\partial\over{\partial t}}
\left({{\vec E}\wedge{\vec H}}\right),\eqno(4)$$
where $n$ is the refractive index of the medium, and $c$ is the velocity
of light {\sl in vacuo}.\par
In the experiment [16] of Walker and Lahoz a disk of barium titanate,
whose relative dielectric constant was near to 4000, had a small cylindrical
hole pierced at its center, and both the inner and the outer cylindrical
surfaces were coated with metallic films. By sending charges on these electrodes
a radial electric field could be set up within the disk. The latter was
suspended to a tungsten wire, and the torsional pendulum obtained in this
way was subjected to a uniform magnetic field of constant value, parallel to
the suspending wire.
When the radial electric field was allowed to vary according to a sine
law at the resonance frequency of the pendulum, steady oscillations were
observed to set in, and their amplitude was measured with an optical lever.
From the known parameters of the experimental device a value for
${\bf{\vec f}}_{exp.}$ was inferred, in good agreement with the
theoretical prediction of eq. [4], as far as can be achieved in a delicate
experiment of this sort.\par
\vbox to 0.6 cm{}

{\gr 3. Reviving a proposal by Brevik}
\bigskip

    It would be interesting to learn whether light exerts Abraham's force
on a transparent medium crossed by it. Since in eq. (4) the product
${\vec E}\wedge{\vec H}$ is differentiated with respect to time, a large
value of the instantaneous force density is predicted in this case, but the
direct observation of a mechanical force at the frequency of light is beyond
reach also for present day experimenters. Brevik [18] made some proposals for
detecting the Abraham's force of light in an indirect way, and one of them will
be reconsidered here. An order of magnitude estimate of several quantities
intervening in the experiment is needed for assessing whether the latter
is actually feasible.\par
    Imagine a single-mode optical fibre {\footnote{\foot $^{5)}$}{For
light with a wavelength in vacuo $\lambda=1.3 \mu m$, as will be chosen
here, a single-mode fibre is typically constituted by a cylindrical doped glass
core with a radius $r_1=5 \mu m$, surrounded by a cylindrical shell of pure
silica glass with a slightly lower refractive index of radius $r_2=60 \mu m$,
and by a further cylindrical coating made of acrylic resin of still lower
refractivity, up to an outer radius $r_3=125 \mu m$. For an account of the
way optical fibres behave see e.g. Refs. [19] and [20].}}, whose length
is $L=1 km$. The mathematical description of the eigenmodes of the
electromagnetic field propagating in such a fibre is rather involved [21],
but if we content ourselves with an order of magnitude evaluation, the actual
field for the so called transverse electric mode can be replaced with the
field of a plane wave propagating along the axis of the fibre, whose
intensity is constant within the fibre core, and vanishes abruptly
at the core radius. Let the axis of the fibre coincide with the $x$ axis
of a Cartesian co-ordinate system, and the electromagnetic plane wave be
a linearly polarized one{\footnote{\foot $^{6)}$}{Polarization preserving optical
fibres are a recent achievement; in usual fibres the polarization is elliptical
with irregular variations of the parameters of the ellipse as one proceeds along
the axis. This circumstance will not affect our order of magnitude estimate.}}
with the nonvanishing components:

$$E_y=E_0cos({k_1}x-{\omega_1}t)cos({k_2}x-{\omega_2}t),\eqno(5)$$
$$B_z={\sqrt{\epsilon_0\epsilon\mu_0\mu}}E_0cos({k_1}x-{\omega_1}t)
cos({k_2}x-{\omega_2}t),\eqno(6)$$
where $\omega_1\ll\omega_2$, and

$${{\omega_1}\over{k_1}}={{\omega_2}\over{k_2}}={c\over n},\eqno(7)$$
{\sl i.e.} we imagine sending{\footnote{\foot $^{7)}$}{A fibre laser [20] can
be directly joined to the fibre without interposition of reflecting surfaces.
Eqs. (5) and (6) do not account for losses; this is a reasonable approximation,
since the attenuation of the fibre considered above can be as low as 0.3 dB,
the limiting value due to Rayleigh scattering.}} into the fibre linearly
polarized light whose amplitude is modulated with the low angular frequency
$\omega_1$. Let us set $\mu=1$, as appropriate for glass. Then Abraham's
force density, defined by eq. (4), is directed along the axis of the fibre
and takes the value:

$${\bf f}=n{{n^2-1}\over{4c}}\epsilon_0E_0^2
{\partial\over{\partial t}}\left\{\left[1+cos2({k_1}x-{\omega_1}t)\right]
\left[1+cos2({k_2}x-{\omega_2}t)\right]\right\}.\eqno(8)$$
The right-hand side of eq. (8) contains a nonvanishing component at low
frequency that is potentially observable; it reads:

$${\bf f}_{low}=n{{n^2-1}\over{2c}}\epsilon_0E_0^2\omega_1
sin2({k_1}x-{\omega_1}t).\eqno(9)$$
The average intensity of the radiation described by eqs. (5) and (6) is
$\bar I={1\over 4}\epsilon_0cE_0^2$. Let us assume for $E_0$ the value
$$E_0=10^6{volt\over m};$$
then the previously quoted size of the core radius allows the fibre to carry
the power of $\sim 60 mW$, an acceptable choice. Imagine now that the fibre
is wound, together with a second one, on a cylinder with radius $R=10 cm$;
this move looks possible without altering in an appreciable way the fields
prevailing in the core. The cylinder is suspended to such a tungsten
wire as to constitute a torsion pendulum whose resonance angular frequency
$\omega$ is

$$\omega=2\omega_1=2\pi s^{-1}.$$
Due to the low frequency component of Abraham's force the first fibre will
exert on the torsion pendulum the torque{\footnote{\foot $^{8)}$}{If both
ends of the fibre are left loose from the torsion pendulum and rigidly
clamped, surface forces occurring at the extremities will not contribute
to the torque.}}, say:

$$M=\pi r_1^2 L{\bf f}_{low} R= M_0sin\omega t,\eqno(10)$$
where, due to the previous choices, and since in the core $n=1.44$ at the
wavelength chosen for the light:

$$M_0=0.63\times 10^{-15} Nm.$$
The deflection angle $\vartheta$ of the torsion pendulum shall
fulfil the equation
$$I\ddot \vartheta+d\dot \vartheta+k\vartheta=M_0sin\omega t,\eqno(11)$$
where $I$, $d$ and $k$ are respectively the momentum of inertia, the damping
parameter and  the elastic constant. For very small damping the resonance
angular frequency $\omega$ is practically equal to the natural frequency of the
undamped pendulum, {\sl i.e.} $\omega=\sqrt{k/I}$. The standing oscillation
at that frequency, according to eq. (11), obeys the law:
$$\vartheta=-{M_0\over{\omega d}}cos\omega t.\eqno(12)$$
We can estimate the damping parameter $d$ by reminding that the time constant
$\tau$ of the damped oscillations is related to $d$ by the equation $\tau=2I/d$.
Since a reasonable value for the time constant is $\tau\approx 10^3 s$, while
$I\approx 2.5\times10^{-3}kg\times m^2$, the standing oscillation at resonance
will occur with the tangential velocity:

$$v=R\dot\vartheta={{M_0 R}\over d}sin\omega t,$$
where one can set

$${{M_0 R}\over d}\approx 1.26\times10^{-11} ms^{-1}.$$

We propose availing of the second fibre to detect this velocity.
We shall disregard the fact that the fibre, whose length can be again
$L=1 km$, is wound on a cylinder, and reason as if it had been uncoiled.
We shall assume that when this fibre is at rest its core, parallel to, say,
the $x$ axis, is travelled by the unmodulated, linearly polarized light
{\footnote{\foot$^{9)}$}{For reasons apparent in the sequel, the second
fibre needs to be polarization preserving.}} whose components, written
{\sl more relativistico}, are:

$$F_{24}=E'_0cos(k_2x-\omega_2t),\eqno(13)$$
$$F_{12}={\sqrt{\epsilon\mu}}E'_0cos(k_2x-\omega_2t).\eqno(14)$$
We imagine however that the velocity field has followed the fibre in the
deformation due to the unwinding action; given the smallness of $v$, we
can approximate the four-velocity of the uncoiled fibre as:

$$u^i\approx(\beta,0,0,1),\eqno(15)$$
where

$$\beta={{M_0 R}\over{cd}}sin\omega t
\approx-{{M_0 R}\over{cd}}sin[2(k_1x-\omega_1 t)],\eqno(16)$$
because $k_1x$ is a negligible term also for $x=L$. Since the fibre is
subjected to this velocity field, eqs. (13) and (14) no longer provide
an exact solution to the Maxwell's equations: the constitutive
relation has changed, however slightly, in keeping with eq. (A14).
Due to the smallness of the change, calculating a first order perturbation
may suffice. If $\delta F^{rs}$ is the change undergone by the unperturbed
$F^{rs}$, the corresponding change in $H^{rs}$ will be given by:
$$\delta H^{rs}={1\over\mu}\delta F^{rs}
+{{\epsilon\mu-1}\over\mu}(\delta^r_4\delta F^{4s}-\delta^s_4\delta F^{4r})
+{{\epsilon\mu-1}\over\mu}\beta(\delta^r_1 F^{4s}-\delta^s_1 F^{4r}
+\delta^r_4 F^{s1}-\delta^s_4 F^{r1}).\eqno(17)$$
Hence in the present case the first order correction to one unperturbed
set of Maxwell's equations reads:
$${1\over\mu}\delta F^{\lambda\rho}_{~,\rho}
+\epsilon\delta F^{\lambda 4}_{~,4}
={{\epsilon\mu-1}\over\mu}
\left[(\beta F^{4\lambda})_{,1}-(\beta F^{1\lambda})_{,4})\right]\eqno(18)$$
for $r=\lambda$, and

$$\epsilon\delta F^{4\rho}_{~,\rho}=0,\eqno(19)$$
for $r=4$, while the correction to the other unperturbed set is:

$$\delta F_{[ik,m]}=0.\eqno(20)$$
Eqs. (18)-(20) display in $\delta F^{ik}$ the same form as the unperturbed
equations do in $F^{ik}$, with the exception of the component with
$\lambda=2$ of eq. (18), that reads:

$$\delta F^{21}_{~,1}+\epsilon\mu\delta F^{24}_{~,4}
=2(\epsilon\mu-1)(\beta F^{42})_{,1}\eqno(21)$$
since in our case:

$$(\beta F^{42})_{,1}=-(\beta F^{12})_{,4}.$$
Therefore only $\delta F^{21}$ and $\delta F^{24}$ need to be nonvanishing.
Let us set:

$$\zeta=(2k_1+k_2)x-(2\omega_1+\omega_2)t,$$
$$\eta=(2k_1-k_2)x-(2\omega_1-\omega_2)t,$$
and

$$C={\textstyle{1\over 2}}(\epsilon\mu-1){M_0RE'_0\over{cd}};$$
then a physically appropriate particular solution is given by:

$$\delta F_{12}=C\left\{sin{\zeta}+sin{\eta}
+(2k_1+k_2)xcos{\zeta}+(2k_1-k_2)xcos{\eta}\right\},\eqno(22)$$
$$\delta F_{42}=-{C\over{\sqrt{\epsilon\mu}}}
\left\{(2k_1+k_2)xcos{\zeta}+(2k_1-k_2)xcos{\eta}\right\}.\eqno(23)$$
We are interested in the behaviour of $\delta F_{ik}$ near the end of
the second fibre, therefore we can retain only the terms that grow
linearly with $x$.
Since $ k_1\ll k_2$, we eventually write:
\vfill\break
$$\delta F_{12}\approx 2\pi(\epsilon\mu-1){M_0RE'_0\over{cd}}
{x\over\lambda_2}sin\omega t~sin(k_2x-\omega_2 t),\eqno (24)$$
$$\delta F_{24}\approx 2\pi{{\epsilon\mu-1}\over\sqrt{\epsilon\mu}}
{M_0RE'_0\over{cd}}{x\over\lambda_2}
sin\omega t~sin(k_2x-\omega_2 t),\eqno (25)$$
where $\lambda_2$ is the wavelength of light {\sl in vitro}. We remark
that $\delta F_{ik}$, at variance with the unperturbed $F_{ik}$, exhibits
a ``sin'' rather than a ``cos'' dependence on the argument $k_2x-\omega_2 t$.
The detection of the signal could be done by adding
{\footnote{\foot $^{10)}$}{Fibre couplers just built with this scope in mind
are commonly available. Of course the whole procedure of detection is
only possible if the second fibre is polarization preserving, and if the
coherence length of the light sent in the second fibre is longer than the
fibre itself. The latter condition ultimately means [20] a request about the
phase and the amplitude stability of the laser source that can be met with
by present day lasers.}}
at the end of the second fibre the field with the nonvanishing components:

$$F_{12}={\sqrt{\epsilon\mu}}E'_0sin(k_2x-\omega_2t).\eqno(26)$$
$$F_{24}=E'_0sin(k_2x-\omega_2t).\eqno(27)$$
Then the square of the electric field $E_y$ has a component whose
low frequency average, that can be perceived by a photodetector, reads:

$$(E_y^2)_{av.}=(E'_0)^2[1-2.\times 10^{-10}sin\omega t].\eqno(28)$$
Although quite small, when compared with the background, the signal caused
by the component of $(E_y^2)_{av.}$ at the resonance frequency of the
torsion pendulum seems within reach of the present day techniques of digital
phase-sensitive detection. Its observation would provide evidence that
Abraham's force is indeed exerted by light inside a dielectric; in fact,
electrostriction [7], [12] in glass should produce a signal that is smaller
in magnitude, and opposite in sign with respect to the one calculated above.
\par
\vbox to 1.0 cm{}

{\gr Acknowledgements}\bigskip

We thank D.-E. Liebscher for his advice and support, and for critically
reading the manuscript. Helpful discussions with M. Allegrini, M. Labardi
and G.C. La Rocca on the issue of the present paper are also gratefully
acknowledged.
\par
\vfill\break

{\gr Appendix: Gordon's derivation of Abraham's tensor}\bigskip

The general relativistic argument conceived by W. Gordon has been retrieved
in the limbo of the forgotten papers [22]. It runs as follows. In a
four-dimensional differentiable manifold let us consider  a  contravariant
skew tensor density ${\bf H}^{ik}$, a covariant skew tensor $F_{ik}$, and
write the naturally invariant equations:

$${\bf H}^{ik}_{  ,k}={\bf s}^i,\eqno(A1)$$

$$F_{[ik,m]}=0.\eqno(A2)$$
The vector density ${\bf s}^i$, defined by the left-hand side of eq. (A1),
represents the electric four-current  density,  while
the comma  signals  ordinary  differentiation,  and  we  have  set
$F_{[ik,m]}\equiv{\textstyle{1\over3}}(F_{ik,m}+F_{km,i}+F_{mi,k})$.
Equations (A1) and (A2) express Maxwell's equations in general curvilinear
co-ordinates; they need to be complemented with the constitutive relation of
electromagnetism, a tensor equation that uniquely defines {\sl e.g.} ${\bf H}^{ik}$
in terms of $F_{ik}$ and of whatever additional  fields  may  be  needed
for  specifying  the properties of the electromagnetic medium. When the
assumed dependence of ${\bf H}^{ik}$ on $F_{ik}$ is algebraic and linear,
the constitutive relation reads:

$${\bf H}^{ik}=\textstyle{1\over2}{\bf X}^{ikmn}F_{mn},\eqno(A3)$$
and the electromagnetic properties of the medium are summarized by
the four-index tensor density ${\bf X}^{ikmn}$. In the case of vacuum
one writes:

$${\bf X}^{ikmn}_{vac.}=\sqrt{g}(g^{im}g^{kn}-g^{in}g^{km});\eqno(A4)$$
the constitutive relation entails only the metric tensor $g_{ik}$ and
$g\equiv-det(g_{ik})$ in the way known from general  relativity. In this
case Maxwell's equations are usually written as:

$${\bf F}^{ik}_{  ,k}={\bf s}^i,\eqno(A5)$$
$$F_{[ik,m]}=0,\eqno(A6)$$
in terms of a skew  tensor $F_{ik}$ and of the contravariant
tensor density ${\bf F}^{ik}\equiv\sqrt{g}g^{im}g^{kn}F_{mn}$.
A general skew tensor $F_{ik}$ can always  be  written as the sum of
the curl of a potential and of the dual to the  curl of an ``antipotential'':

$$F_{ik}=\varphi_{k,i}-\varphi_{i,k}
+e_{ik}^{~~mn}(\psi_{n,m}-\psi_{m,n}),\eqno(A7)$$
where $e_{ik}^{~~mn}$  is the tensor obtained from the  Ricci-Levi Civita
symbol ${\bf e}^{ikmn}$ in  the  usual  way. Starting from the Lagrangian
density

$${\bf L}=\textstyle{1\over4}{\bf F}^{ik}F_{ik}-{\bf s}^i\varphi_i\eqno(A8)$$
for the general relativistic vacuum, both sets of Maxwell's equations can be
derived through  the  Hamilton principle, by  asking  that  the variations
of $A=\int{\bf L}\,dS$  ($\,dS=\,dx^1\,dx^2\,dx^3\,dx^4$) with respect to
$\varphi_i$  and to $\psi_i$ separately vanish [23].
The well known form of the energy tensor density ${\bf T}_{ik}$
for  the electromagnetic field {\sl in vacuo} is obtained
through  the method [24] inaugurated by Hilbert, {\sl i.e.} by carrying out
the Hamiltonian derivative of the Lagrangian  density (A8) with respect to
$g^{ik}$:

$${\bf T}_{ik}\equiv 2{\delta{\bf L}\over\delta{g^{ik}}}
={\bf F}_i^{~n}F_{kn}-\textstyle{1\over4}
g_{ik}{\bf F}^{mn}F_{mn}.\eqno(A9)$$
The  derivation  of  the energy tensor must be performed by keeping into
account  the homogeneous field equations obtained above; they dictate  that
$F_{ik}$ is the curl of a four-vector  $\varphi_i$  and does not contain
the metric.\par

The constitutive relation for a  linear, nondispersive medium will
have in general the form of eq.~(A3). Gordon deals with a medium
that in addition is homogeneous and isotropic, when considered at rest;
its constitutive  relation can be written in a simple form, due to
Minkowski [3], that can be extended without change to general relativity. Let

$$u^i={dx^i\over\sqrt{-ds^2}}\eqno(A10)$$
be the four-velocity of matter, for which $u_iu^i=-1$.  One  defines
the four-vectors

$$F_i=F_{ik}u^k,\quad H_i=H_{ik}u^k,\eqno(A11)$$
where $H_{ik}\equiv{(1/\sqrt{g})}g_{ip}g_{kq}{\bf H}^{pq}$ is
the covariant tensor associated with ${\bf H}^{ik}$. Then the above
mentioned constitutive relation simply reads:

$$H_i=\epsilon{F_i},\eqno(A12)$$
$$u_iF_{km}+u_kF_{mi}+u_mF_{ik}=
\mu\big(u_iH_{km}+u_kH_{mi}+u_mH_{ik}\big),\eqno(A13)$$
where $\epsilon$ and $\mu$ are the dielectric constant and the
permeability, and do not depend on the chosen event. These eight
equations,  that  entail  two identities, are equivalent [25]
to the six equations:

$$\mu{H^{ik}}=F^{ik}+(\epsilon\mu-1)(u^iF^k-u^kF^i).\eqno(A14)$$
Since the right-hand side of eq. (A14) can be rewritten as

$$F_{rs}\left\{g^{ir}g^{ks}
-(\epsilon\mu-1)\left(u^iu^rg^{ks}+u^ku^sg^{ir}\right)\right\},$$
and,  due  to  the  antisymmetry of $F_{rs}$, one can freely add the
term $(\epsilon\mu-1)^2u^iu^ku^ru^s$  within  the curly brackets, the
constitutive relation eventually comes to read:

$$\mu{H^{ik}}=\left[g^{ir}-(\epsilon\mu-1)u^iu^r\right]
    \left[g^{ks}-(\epsilon\mu-1)u^ku^s\right]F_{rs}.\eqno(A15)$$
The ``effective metric tensor'':

$$\gamma^{ik}=g^{ik}-(\epsilon\mu-1)u^iu^k,\eqno(A16)$$
whose inverse is

$$\gamma_{ik}=g_{ik}+\big(1-{1\over{\epsilon\mu}}\big)u_iu_k,\eqno(A17)$$
allows one to write eq. (A14) in the form:

$$\mu{\bf H}^{ik}=\sqrt{g}\gamma^{ir}\gamma^{ks}F_{rs}.\eqno(A18)$$
Since $g\equiv-det(g_{ik})$, we shall pose $\gamma\equiv-det(\gamma_{ik})$.
The  ratio $\gamma/g$ is an invariant, and its calculation
can be performed in the co-ordinate system in which $u^1=u^2=u^3=0$;
one finds:

$$\gamma={g\over{\epsilon\mu}},$$
so that eq.~(A15) can be rewritten as

$${\bf H}^{ik}=\sqrt{\epsilon\over\mu}
\sqrt{\gamma}\gamma^{ir}\gamma^{ks}F_{rs},\eqno(A19)$$
which, {\sl apart from  the  constant  factor} $\sqrt{\epsilon/\mu}$,
is  just  the constitutive relation for the vacuum case in general
relativity, when $\gamma_{ik}$ acts as metric.\par

We shall henceforth enclose in  round  brackets  the  indices
which  are  either  moved  with $\gamma^{ik}$  and $\gamma_{ik}$,  or
generated  by performing a Hamiltonian derivative with respect to
the  latter tensors; therefore eq.~(A19) will be rewritten as

$${\bf H}^{ik}=\sqrt{\epsilon\over\mu}
\sqrt{\gamma}F^{(i)(k)}.\eqno(A20)$$
Due to the strict analogy with the vacuum case made evident in this way,
selecting the form of the Lagrangian density  for  the  electromagnetic
field in the medium under question is for Gordon a straightforward matter.
He writes:

$${\bf L}'={\textstyle{1\over4}}\sqrt{\epsilon\over\mu}
\sqrt{\gamma}F^{(i)(k)}F_{ik}-{\bf s}^i\varphi_i,\eqno(A21)$$
where $F_{ik}$ can be defined by:

$$F_{ik}=\varphi_{k,i}-\varphi_{i,k}+{1\over{\sqrt\gamma}}
{\bf e}_{(i)(k)}^{~~~~~~mn}(\psi_{n,m}-\psi_{m,n}).\eqno(A22)$$
Equating to  zero the independent variations of $\int{{\bf L}'\,dS}$ with
respect to $\varphi_i$ and to $\psi_i$ will produce Maxwell's equations,
{\sl a priori} complemented by the constitutive  relation (A14).\par
It is now easy to  derive  the  energy  tensor  for   the
electromagnetic field by starting from  the  derivation  that  one
performs {\sl in vacuo}, when the metric field $\gamma_{ik}$ is present.
In that case Hilbert's procedure leads to write:

$$\delta{\bf L}\equiv{\textstyle{1\over2}}
{\bf T}_{(i)(k)}\delta\gamma^{ik},$$
hence one gets

$${\bf T}_{(i)}^{~~(k)}=\sqrt\gamma\big(F_{ir}F^{(k)(r)}
-\textstyle{1\over4}\delta_i^{~k}F_{rs}F^{(r)(s)}\big).\eqno(A23)$$
For the medium contemplated by Gordon one may be tempted to write:

$$\delta{\bf L}'\equiv{\textstyle{1\over2}}
{\bf T}'_{(i)(k)}\delta\gamma^{ik},\eqno(A24)$$
where ${\bf L}'$ is given by eq. (A21), and finds

$${\bf T}_{(i)}^{'~~(k)}=F_{ir}{\bf H}^{kr}
-\textstyle{1\over4}\delta_i^{~k}F_{rs}{\bf H}^{rs},\eqno(A25)$$
which is  just  the  general  relativistic  version  of  the  form
proposed by Minkowski. Henceforth we  shall  drop the prime, since
its omission will not lead to confusion.\par
However, ${\bf T}_{(i)}^{~~(k)}$ can not be the energy tensor density
that  we are looking for, because it is defined with respect to  the
effective metric  $\gamma_{ik}$, not with respect to the true metric
$g_{ik}$, the only  one entitled to account  for  the structure of space-time
and, via  the Einstein tensor, for its stress-energy-momentum tensor.
In order to find the relation  between ${\bf T}_{ik}$ and
${\bf T}_{(i)(k)}$ we simply need to express $\delta\gamma^{ik}$
in terms of $\delta g^{ik}$. From  eq. (A10) one obtains the variation
of $u^i$  produced by the variation $\delta g^{mn}$ of  the metric:

$$\eqalign\delta u^i={\textstyle{1\over2}}u^iu^mu^n\delta g_{mn}
=-{\textstyle{1\over2}}u^iu_mu_n\delta g^{mn},$$
hence from the definition (A16) one gets:

$$\eqalign\delta\gamma^{ik}=\delta\big\{g^{ik}-(\epsilon\mu-1)u^iu^k\big\}
=\delta g^{ik}+(\epsilon\mu-1)u^iu^ku_mu_n\delta g^{mn}.\eqno(A26)$$
Therefore, since ${\bf T}_{ik}\delta g^{ik}={\bf T}_{(i)(k)}\delta\gamma^{ik}$,
we find immediately:

$$\eqalign{\bf T}_{ik}={\bf T}_{(i)(k)}
+(\epsilon\mu-1)u_iu_k{\bf T}_{(m)(n)}u^mu^n.\eqno(A27)$$

The mixed components of ${\bf T}_{ik}$   can be  obtained  by  multiplying
the left-hand side and the second term at the right-hand  side  of
eq.~(A27) by $g^{kq}$, while the first term at the  right-hand  side  is
multiplied by ${\gamma^{kq}+(\epsilon\mu-1)u^ku^q}$.
Then

$${\bf T}_i^{~q}={\bf T}_{(i)}^{~~(q)}
+(\epsilon\mu-1)\big\{{\bf T}_{(i)(k)}u^k
+u_i{\bf T}_{(m)(n)}u^mu^n\big\}u^q.\eqno(A28)$$
In a co-ordinate system for which, at a given event,
$g_{ik}=\eta_{ik}\equiv{diag(1,1,1,-1)}$ and
$u^1=u^2=u^3=0,\ u^4=1$, the covariant four-vector within the curly
brackets has the components ${\bf T}_{(\alpha)(4)},\ 0\  (\alpha=1,2,3)$.
But under  these circumstances eq.~(A27) says that
${\bf T}_{(\alpha)(4)}={\bf T}_{\alpha4}$,  hence in a general
co-ordinate system one can write:

$${\bf T}_i^{~q}={\bf T}_{(i)}^{~~(q)}
+(\epsilon\mu-1)\big\{{\bf T}_{ik}u^k
+u_i{\bf T}_{mn}u^mu^n\big\}u^q,\eqno(A29)$$
{\sl i.e.}, according to eq. (A25)

$$T_i^{~k}=F_{ir}H^{kr}-\textstyle{1\over4}
\delta_i^{~k}F_{rs}H^{rs}-(\epsilon\mu-1)\Omega_iu^k,\eqno(A30)$$
where Minkowski's ``Ruh-Strahl''

$$\Omega^i=-\big(T_n^{~i}u^n+u^iT_{mn}u^mu^n\big)\eqno(A31)$$
has been introduced. Since $\Omega^iu_i=0$, by substituting (A30) into (A31)
one finds:

$$\eqalign\Omega^i=F_mH^{im}-F_mH^mu^i=u_kF_m
\big(H^{ik}u^m+H^{km}u^i+H^{mi}u^k\big).\eqno(A32)$$
Equations (A30) and (A32) define the extension to general relativity
of the energy tensor proposed by  Abraham [4], [5]  for  the
electromagnetic field if the medium is homogeneous and isotropic
when considered at rest.\par
The generalized force density due to the electromagnetic field is
eventually defined by the covariant divergence:
$${{\bf f}_i}\equiv{-{\bf T}^{~k}_{i~;k}}.\eqno(A33)$$
\par
\vbox to 1.0 cm{}

{\gr References}\par\bigskip

[1]~~{\nom Helmholtz, H.}, {\sl Ann. d. Physik u. Chemie}, {\bf13}, 385 (1881).

[2]~~{\nom v. Helmholtz, H.}, {\sl Ann. d. Physik u. Chemie}, {\bf47}, 1 (1892).

[3]~~{\nom Minkowski, H.}, G\"ott. Nachr., Math.-phys. Klasse (1908), 53.

[4]~~{\nom Abraham, M.}, Rend. Circ. Matem. Palermo {\bf 28}, 1 (1909).

[5]~~{\nom Abraham, M.}, Rend. Circ. Matem. Palermo {\bf 30}, 33 (1910).

[6]~~{\nom Einstein, A. and Laub, J.}, {\sl Ann. d. Physik}, {\bf26}, 541 (1908).

[7]~~{\nom Robinson, F.N.H.}, {\sl Phys. Reports}, {\bf16}, 313 (1975).

[8]~~{\nom Israel, W.}, {\sl Gen. Rel. Grav.}, {\bf9}, 451 (1978).

[9]~~{\nom Krany\v s, M.}, {\sl Can. J. Phys.}, {\bf57}, 1022 (1979).

[10]~{\nom Maugin, G.A.}, {\sl Can. J. Phys.}, {\bf58}, 1163 (1980).

[11]~{\nom Synge, J.L.}, {\sl Hermathena}, {\bf117}, 80 (1974).

[12]~{\nom Brevik, I.}, {\sl Phys. Reports}, {\bf52}, 133 (1979).

[13]~{\nom Kristensen, M. and Woerdman, J.P.}, {\sl Phys. Rev. Lett},
{\bf72}, 2171 (1994).

[14]~{\nom Gordon, W.}, {\sl Ann. d. Physik}, {\bf72}, 421 (1923).

[15]~{\nom James, R.P.}, {\sl Proc. Nat. Acad. Sci.}, {\bf61}, 1149 (1968).

[16]~{\nom Walker, G.B. and Lahoz, D.G.}, {\sl Nature}, {\bf253}, 339 (1975).

[17]~{\nom Walker, G.B., Lahoz, D.G. and Walker, G.}, {\sl Can. J. Phys.},
{\bf53}, 2577\par
~~~~~(1975).

[18]~Ref. [12], pp. 188-191.

[19]~{\nom Adams, M.J.}, {\sl An Introduction to Optical Waveguides},
(J. Wiley) 1981.

[20]~{\nom Agrawal, G.P.}, {\sl Nonlinear Fiber Optics}, (Academic Press) 1995.

[21]~see e.g. Ref. [19], Ch. 7.

[22]~{\nom Antoci, S. and Mihich, L.}, Nuovo Cimento B, {\bf 112}, 991 (1997).

[23]~{\nom Finzi, B.}, Rend. Accad. Naz. Lincei {\bf 12}, 378 (1952).

[24]~{\nom Hilbert,~D.}, G\"ott. Nachr., Math.-phys. Klasse (1915), 395.

[25]~{\nom Nordstr\"om, G.}, Soc. Scient. Fenn., Comm. Phys.-Math. {\bf 1.33},
1 (1923).
\end